\documentclass[a4paper,11pt,english,oneside]{article}
\pdfoutput=1

\usepackage{geometry}
\usepackage{amsmath}
\usepackage{amssymb}
\usepackage{color}
\usepackage{graphicx}
\usepackage{subfig}
\usepackage{booktabs}
\usepackage{verbatim}
\usepackage{float}
\usepackage[]{units}
\usepackage{enumitem}
\usepackage[mathcal]{euscript}
\DeclareMathAlphabet{\mathpzc}{OT1}{pzc}{m}{it}

\usepackage[T1]{fontenc}
\usepackage[latin9]{inputenc}
\usepackage{float}
\usepackage{babel}

\providecommand{\tabularnewline}{\\}

\newcommand{\TeV}{\,\mathrm{TeV}}
\newcommand{\GeV}{\,\mathrm{GeV}}
\newcommand{\MeV}{\,\mathrm{MeV}}

\newcommand{\fracwithdelims}[4]{\Bigl#1 \frac{#3}{#4} \Bigr#2}
\newcommand{\ord}[1]{\mathcal{O}\left( #1 \right)}

\newcommand{\vev}[1]{\left\langle #1\right\rangle}
\newcommand{\VeV}[2]{#1\langle #2 #1\rangle} 

\newcommand{\Fig}[1]{Fig.~\ref{fig:#1}}

\newcommand{\eq}[1]{eq.~(\ref{eq:#1})}
\newcommand{\Eqs}[1]{Eqs.~(\ref{eq:#1})}
\newcommand{\eqs}[1]{eqs.~(\ref{eq:#1})}

\newcommand{\nohyphens}%
       {\hyphenpenalty=10000\exhyphenpenalty=10000\relax}

\DeclareMathOperator{\tr}{Tr}

\newcommand{\GSM}{G_\text{SM}}
\newcommand{\Uc}{U^c}
\newcommand{\Dc}{D^c}
\newcommand{\Qc}{Q^c}
\newcommand{\g}{\hat g}

\newcommand{\SU}[1]{\mathcal{SU}(#1)}
\newcommand{\gen}[1]{\mathpzc{t}_{#1}}
\newcommand{\Gen}[1]{\mathpzc t_{#1}}
\newcommand{\gh}[1]{\mathpzc g_{#1}}

\allowdisplaybreaks[1]

\newlength{\myem}
\newcounter{mysubequation}[equation]

\newcommand{\SISSA}{SISSA/ISAS and INFN, I--34151 Trieste, Italy}

\newcommand{\preprintnumber}{%
SISSA--20/2012/EP}

\newcommand{\titletext}{Simple and direct communication of dynamical supersymmetry breaking}
\newcommand{\authortext}{\large Francesco Caracciolo and Andrea Romanino
\medskip\\\em\normalsize
\SISSA
}
\newcommand{\abstracttext}{We present a complete, calculable, and phenomenologically viable model of dynamical supersymmetry breaking. The model is a simple extension of the so called 3--2 model, with gauge group $\SU{3}\times \SU{2} \times G_\text{SM}$ and the MSSM fields directly coupled to the hidden sector $\SU{2}$ vector fields. Sfermion masses are universal, thus solving the supersymmetric flavour problem, and gaugino masses are not suppressed, in fact they are predicted to be of the same order as sfermion masses. Sizeable contributions to the MSSM $A$-terms can be generated, depending on the size of some free couplings.
As a byproduct, we show some properties of a class of models with $n$ pairs of Higgs doublets.}

\title{
\normalsize
\hspace*{\fill}
\begin{tabular}[t]{l}\preprintnumber\end{tabular}
\vspace{3\baselineskip}\\
\Large\bfseries\titletext\bigskip}
\author{\begin{minipage}[t]{0.8\textwidth}
\normalsize\centering\authortext
\end{minipage}}
\date{}

\begin{document}

\bigskip
\maketitle
\begin{abstract}\normalsize\noindent
\abstracttext
\end{abstract}\normalsize\vspace{\baselineskip}





\section{Introduction}

If supersymmetry is realized in Nature, it has to be broken. From an aesthetical point of view, models in which supersymmetry is broken spontaneously and dynamically~\cite{Witten:1981nf} are particularly appealing. From a phenomenological point of view, we need the sfermion mass terms to be flavour universal, at least in the first two families, and, if the naturalness criterium is not abandoned~\cite{ArkaniHamed:2004fb}, gaugino masses to be roughly of the same order of magnitude as sfermion masses.

Gauge-mediation models~\cite{Dine:1981za} satisfy the flavour constraint. Viable gaugino masses can also be obtained if supersymmetry breaking is parameterized by a spurion field, as in minimal gauge mediation~\cite{Dimopoulos:1996vz}. On the other hand, when a concrete supersymmetry breaking sector is incorporated, gauge mediation models sometimes fail to provide large enough gaugino masses. Indeed, gaugino masses seem to represent an obstacle to obtain a phenomenologically viable model of dynamical supersymmetry breaking. One reason has to do with the R-symmetry. If an R-symmetry is present (which is the case in generic models with stable supersymmetry breaking minima~\cite{Nelson:1993nf}), it needs to be broken in order for non-vanishing (Majorana) gaugino masses to be allowed\footnote{The possibility of Dirac gaugino masses is studied in~\cite{Fayet:1978qc}).}. On the other hand, dynamical models often flow at low energy to generalized O'Raifeartaigh models with R-charges 0 and 2 (see however~\cite{Bertolini:2011hy}), in which the R-symmetry might not be spontaneously broken~\cite{Shih:2007av}. Even when the R-symmetry is spontaneously broken, gaugino masses can turn out to be strongly suppressed, as in semi-direct Gauge Mediation~\cite{Seiberg:2008qj,ArkaniHamed:1998kj}. Independently of the R-symmetry, gaugino masses turn out to vanish at the one loop if the dynamical model has a generalized O'Raifeartaigh low-energy limit in which the supersymmetry breaking pseudoflat direction is stable everywhere~\cite{Komargodski:2009jf}.

\begin{figure}
 \centering
 \includegraphics[width=0.5\textwidth]{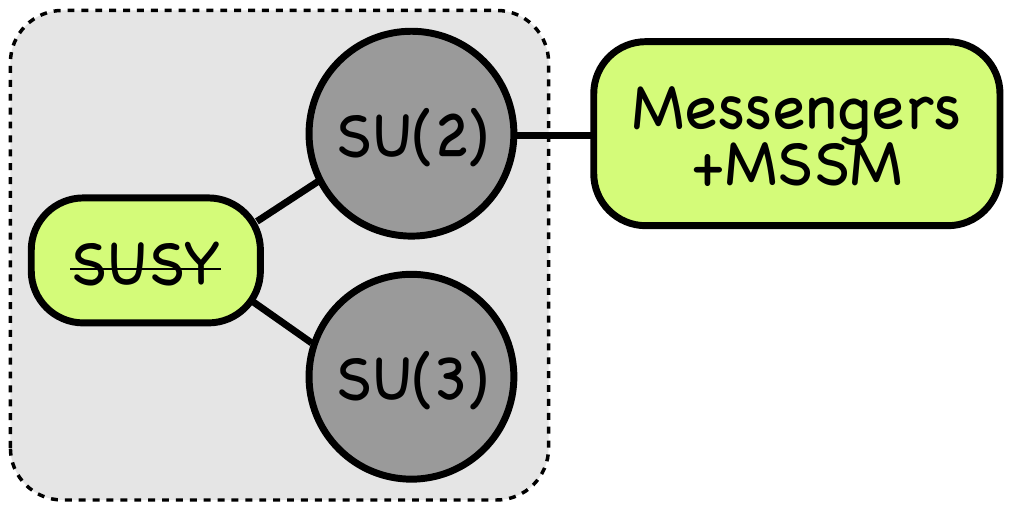}
 \caption{Schematic representation of the model. Supersymmetry breaking takes place in the the 3--2 sector (shaded rectangle). The supersymmetry breaking fields are charged under the $\SU{3}\times\SU{2}$ gauge group. The MSSM fields also feel the $\SU{2}$ interactions, which communicate supersymmetry breaking to the sfermions at the tree level. The MSSM fields are unified in $\SU{2}$ doublets with heavy fields (behaving like the messengers of minimal gauge mediation). The latter get their mass from a superpotential coupling to the source of $\SU{2}$ and supersymmetry breaking in the 3--2 sector. }
\label{fig:32}
\end{figure}

In this paper we present a simple, phenomenologically viable model of dynamical supersymmetry breaking providing universal sfermion masses and non vanishing gaugino masses of the same order. The model is a simple extension of the 3--2 model of dynamical supersymmetry breaking~\cite{Affleck:1984xz}, with gauge group $\SU{3}\times \SU{2} \times G_\text{SM}$, where $G_\text{SM} = \text{SU(3)}_c\times\text{SU(2)}_L\times \text{U(1)}_Y$ is the SM gauge group. Some of the features of the model are:
\begin{itemize}
\item
Unlike in ordinary gauge-mediation, supersymmetry breaking is communicated to the MSSM fields by the $\SU{2}$ gauge interactions, not by SM gauge interactions.
\item
The MSSM fields are unified in $\SU{2}$ doublets with heavy fields playing the role of the chiral messenger of minimal gauge mediation.
\item
The messenger \emph{and} observable fields are directly coupled to the hidden sector $\SU{2}$ gauge fields and to the source of supersymmetry (and $\SU{2}$) breaking.
\item
Both messenger and observable fields are charged under the hidden sector gauge group (the weak part $\SU{2}$), but they do not take part to supersymmetry breaking.
\item
The messenger + observable sector is analogous to the messenger sector of semi-direct Gauge Mediation, however
\begin{description}[leftmargin=0mm]
\item[-]\hspace*{-2mm}
no additional separate sector for the MSSM fields is required;
\item[-]\hspace*{-2mm}
no explicit mass term is needed for the messengers, which get their masses by coupling to the $\SU{2}$ breaking sources in the 3--2 sector;
\item[-]\hspace*{-2mm}
gaugino masses are not suppressed, they arise at the one loop level because of the above coupling of the messengers to the supersymmetry breaking source.
\end{description}
\item
Positive sfermion masses arise at the tree level, in what can be considered as a dynamical realization of tree-level gauge mediation (TGM)~\cite{Nardecchia:2009ew}, but are predicted not to be hierarchically larger than the gaugino masses.
\end{itemize}
A schematic representation of the supersymmetry breaking scheme is given in \Fig{32}.

The paper is organized as follows. In Section~\ref{sec:32}, we briefly review the 3--2 model. Section~\ref{sec:coupling} is the core of the paper. There, we show how the supersymmetry breaking originating in the 3--2 sector can be simply communicated to the observable fields. In Section~\ref{sec:yukawas}, the MSSM Higgs and Yukawas are introduced. In Section~\ref{sec:Higgs}, we address mode-dependent issues about the Higgs sector and in Sections~\ref{sec:1loop} and~\ref{sec:2loop} we show that sizeable $A$-terms can arise due to matter-messenger couplings and discuss loop corrections to the sfermion masses. We summarize our results in Section~\ref{sec:conclusions}. In the Appendix, we show some useful results on electroweak symmetry breaking in the presence of $n$ pairs of Higgs doublets.

\section{The 3--2 sector}
\label{sec:32}

Let us begin by reviewing the 3--2 model. The gauge group is $\SU{3}\times \SU{2}$ (we use calligraphic letters for the 3--2 groups, generators, and couplings), and the matter content is
\begin{table}[H]
\begin{centering}
\begin{tabular}{ccc}
 & $\SU{3}$ & $\SU{2}$ \tabularnewline
$Q$ & $3$ & $2$\tabularnewline
$\Uc$ & $\overline{3}$ & $1$\tabularnewline
$\Dc$ & $\overline{3}$ & $1$\tabularnewline
$L$ & $1$ & $\overline{2}$\tabularnewline
\end{tabular} .
\par\end{centering}
\label{tab:32}
\end{table}
\noindent
For convenience, we assume $L$ to transform as $L\to U^* L$ under $U\in\SU{2}$. The superpotential $W_\text{3--2}$ of the 3--2 model is the sum of two terms:
\begin{equation}
\label{eq:W32}
W_\text{3--2} = W_\text{cl} + W_\text{np} .
\end{equation}
The classical term is
\begin{equation}
W_\text{cl}=hQ\Dc L\equiv hQ_{A}^{a}\Dc_aL^{A},
\label{eq:1-1}
\end{equation}
where capital letters correspond to $\SU{2}$ indices, lower-case letters to $\SU{3}$ indices. The $\SU{3}$ interactions become non-perturbative at the scale $\Lambda_3$, giving rise to the term
\[
W_\text{np}=\frac{\Lambda_3^{7}}{\text{det}Q\Qc}, \qquad
\Qc \equiv (\Dc, \Uc), \qquad
\text{det}Q\Qc=Q_{A}^{r}Q_{B}^{s}\Dc_{r}\Uc_{s}\epsilon^{AB}.
\]
The $\SU{2}$ interactions are assumed to be perturbative at that scale (and above). In the further assumption that $h\ll 1$ and $h\ll \gh{2},\gh{3}$, the $F$-term contribution to the potential is subleading and the minimum can be obtained perturbatively along the $D$-flat directions. In an appropriate flavour basis,
\begin{equation}
\label{eq:23vevs}
Q=\left(\begin{array}{cc}
a & 0\\
0 & b\\
0 & 0
\end{array}\right) M
\qquad
\Qc=\left(\begin{array}{cc}
a & 0\\
0 & b\\
0 & 0
\end{array}\right) M
\qquad
L=\left(\begin{array}{c}
\sqrt{a^{2}-b^{2}} \\
0
\end{array}\right) M ,
\end{equation}
where
\[
M \equiv \frac{\Lambda_3}{h^{1/7}} \gg \Lambda_3 ,
\]
and $a\approx 1.164$, $b\approx 1.132$. Note that the component of $L$ getting a non-vanishing vev, $L_1$, has $\gen{3} = -1/2$. The $\SU{3}\times \SU{2}$ symmetry is thus fully broken. The $F-$terms are
\begin{equation}
\label{eq:23fterms}
F_{Q} = F_{\Qc} = \left(\begin{array}{cc}
a\sqrt{a^2-b^2}-1/(a^{3}b^{2}) & 0\\
0 & 1/(a^{2}b^{3}) \\
0 & 0
\end{array}\right) F,
\qquad
F_{L}= \left(\begin{array}{c}
a^2 \\
0
\end{array}\right) F,
\end{equation}
where
\[
F \equiv h M^2 = h^{5/7} \Lambda^2_3 \ll \Lambda^2_3 \ll M^2.
\]
The $F$-terms above induce two non-vanishing $D$-terms: $D^{(2)}_3$, associated to the $\Gen{3} = \sigma_3/2$ generator of $\SU{2}$ and $D^{(3)}_3$, associated to the corresponding $\SU{3}$ generator $\lambda_3/2$($\sigma_a$ and $\lambda_A$ are the Pauli and Gell-Mann matrices respectively). The can both be obtained using the general result
\begin{equation}
\label{eq:DvsF}
\vev{D_A} = 2(M^2_V)^{-1}_{AB} \,g_B f^\dagger_0 T_B f_0 ,
\end{equation}
where $f_0$ groups all the $F$-terms, $M^2_V$ is the heavy gauge boson mass matrix, $T_{A,B}$ are broken generators and $g_{A,B}$ the corresponding gauge couplings. We are interested in the $\SU{2}$ $D$-term, which turns out to be
\begin{equation}
\label{eq:D23}
\gh{2} \VeV{}{D^{(2)}_3} = -2c\frac{F^2}{M^2} \text{, with } c =
\frac{2 a^8 b^8+2 a^2+4 a^4 b^4 \sqrt{a^2-b^2}-2 b^2}{3 a^8 b^6-a^6 b^8} \approx 1.48 .
\end{equation}

\bigskip

We now extend the 3--2 model and couple it to the MSSM fields. As anticipated, supersymmetry breaking will be communicated to the MSSM fields by $\SU{2}$ gauge interactions. We will in fact identify the MSSM superfields with the $\Gen{3} = -1/2$ components $f$ of a set of $\SU{2}$ doublets $\Phi = (\phi, f)^T$. Sfermion masses then arise at the tree-level directly from the $\SU{2}$ $D$-term in \eq{D23}:
\begin{equation}
\label{eq:mmphi}
\gh{2}\, \Phi^\dagger \frac{\sigma_a}{2}\Phi \VeV{}{D^{(2)}_a} \rightarrow \tilde m^2_f = - \frac{\gh{2}}{2}\VeV{}{D^{(2)}_3} = c\frac{F^2}{M^2}> 0 .
\end{equation}

\section{Coupling the 3--2 model to the MSSM}
\label{sec:coupling}

Let us extend the 3--2 model by adding the SM gauge factor $\GSM$, a set of $\SU{2}$ doublets $\Phi$ with $\GSM$ quantum numbers corresponding to the three families of SM fermions ($R_\text{SM}$) and three singlets $\nu^c_i$, and a set of $\SU{2}$ singlets with opposite SM quantum numbers.

\begin{table}[H]
\begin{centering}
\begin{tabular}{ccc|c}
 & $\SU{3}$ & $\SU{2}$ & $\GSM$ \tabularnewline
$Q$ & $3$ & $2$ & 1 \tabularnewline
$\Uc$ & $\overline{3}$ & $1$ & 1 \tabularnewline
$\Dc$ & $\overline{3}$ & $1$ & 1 \tabularnewline
$L$ & $1$ & $\overline{2}$ & 1 \tabularnewline
 \hline
$\Phi$ & 1 & 2 & $R_\text{SM+$\nu^c$}$ \tabularnewline
$\overline{\phi}$ & 1 & 1 & $\overline{R}_\text{SM+$\nu^c$}$ \tabularnewline
\end{tabular} .
\par\end{centering}
\end{table}

The doublets $\Phi = (\phi, f)^T$ contain two copies of the MSSM matter fields: $q_i$, $u^c_i$, $d^c_i$, $l_i$, $e^c_i$, $\nu^c_i$ ($i=1,2,3$ is the family index), collectively denoted as $f_\alpha$, $\alpha = 1\ldots 18$, and $\phi^q_i$, $\phi^{u^c}_i$, $\phi^{d^c}_i$, $\phi^l_i$, $\phi^{e^c}_i$, $\phi^{\nu^c}_i$, collectively denoted as $\phi_\alpha$. The second copy $\phi_\alpha$ will get a heavy mass (proportional to $M$) together with $\bar{\phi}_\alpha
 $ through $\SU{2}$ breaking, while the first copy will be massless before electroweak symmetry breaking and will be identified with the MSSM matter superfields. We have included the extra SM singlets $\Phi^{\nu^c}_i = (\nu^c_i, \phi^{\nu^c}_i)$ in order to cancel the $\SU{2}$ Witten anomaly~\cite{Witten:1982fp}.

In the presence of the fields $\Phi_\alpha$, $\overline{\phi}_\alpha$, a new term
\begin{equation}
\label{eq:WM}
W_{M} = y_\alpha L \Phi^{\phantom{\dagger}}_\alpha \overline{\phi}_\alpha = y_\alpha L_1 \phi_\alpha \overline{\phi}_\alpha + y_\alpha L_2 f_\alpha \overline{\phi}_\alpha
\end{equation}
(in an appropriate basis in flavour space) can be added to the 3--2 superpotential $W_\text{3--2}$. 
Assuming as usual $R$-parity conservation to avoid exceedingly large lepton- and baryon-number violating operators, that is the only additional term allowed in the superpotential at the renormalizable level, besides the singlet mass terms $(M^N_{ij}/2) \Phi^{\nu^c}_i\Phi^{\nu^c}_j$ and $(M^{\overline{n}}_{ij}/2) \bar{\phi}^{\nu^c}_i \bar{\phi}^{\nu^c}_j$, which will not play a role in what follows and will therefore be ignored\footnote{Such terms can be used to make the spare SM singlets $\nu^c_i$ heavy and may play a role in generating neutrino masses.}.

We have checked that the introduction of the new fields $\Phi,\overline{\phi}$ and their superpotential (and of the Higgs fields and $\GSM$ gauge interactions) does not destabilize the 3--2 vacuum in \eqs{23vevs} neither at the tree level nor at the one loop level. In particular, the upper component $L_1$ of the $\SU{2}$ doublet $L = (L_1,L_2)^T$ still gets a vev in both the scalar and $F$-term components, $\vev{L_1} = \sqrt{a^2-b^2} M + a^2 F\theta^2$. The $\SU{2}$ breaking vev of the scalar component generates the superpotential mass terms
\begin{equation}
\label{eq:M}
M^{\phantom{\dagger}}_\alpha \phi_\alpha \overline{\phi}_\alpha, \qquad M_\alpha = y_\alpha \sqrt{a^2-b^2} M,
\end{equation}
leaving only the MSSM matter fields $f_\alpha = q_i,u^{c}_i,d^{c}_i,l_i,e^{c}_i$ (and possibly $\nu^c_i$) at the electroweak scale. Moreover, because of the superpotential coupling
\begin{equation}
\label{eq:chiralmessengers}
y_\alpha L_1 \phi_\alpha\overline{\phi}_\alpha ,
\end{equation}
$\phi$ and $\overline{\phi}$ play the role of the chiral messengers of minimal gauge mediation, with $L_1$ playing the role of the spurion field.

Below the scale $M$, the model we have considered so far reduces to the matter and gauge sector of the MSSM, very weakly coupled to the 3--2 fields\footnote{The chiral degrees of freedom of the 3--2 model that are not eaten by the 3--2 gauge superfields get mass at the scale $\tilde m$ or below~\cite{Affleck:1984xz}. In the effective theory below the scale $M$, they are coupled to the observable fields by non-renormalizable operators suppressed by the scale $M$. The light fermionic degrees of freedom of the 3--2 model are three Weyl fermions, one with mass of order $\tilde m$, one with mass possibly induced by higher dimension operators, and the Goldstino, which is eaten by a gravitino with mass $m_{3/2} \sim \tilde m (M/M_\text{Pl})$. The light scalar degrees of freedom are three real and a complex scalar with mass of order $\tilde m$ and the $R$-axion that, in our case, has mass $m^2_R \sim \tilde m^2 (M/M_\text{Pl})$~\cite{Bagger:1994hh,Seiberg:2008qj}.}. Higgs and Yukawa interactions will be introduced in the next subsection. As anticipated, sfermion masses arise directly from the $\SU{2}$ $D$-term in \eq{D23} and are given, at the tree level, by the universal value
\begin{equation}
\label{eq:mm}
\tilde m^2 = - \frac{\gh{2}}{2}\VeV{}{D^{(2)}_3} = c \, \frac{F^2}{M^2} \approx 1.48 \, \frac{F^2}{M^2}.
\end{equation}
Note that sfermion masses arise because supersymmetry breaking is communicated to the MSSM fields by the $\SU{2}$ gauge interactions, not by the SM gauge interactions. Note also that sfermion masses turn out to be
\begin{itemize}
\item
flavour-universal, thus solving the supersymmetric flavour problem;
\item
universal within each family, thus providing a rationale for the CMSSM;
\item
positive, despite they arise at the tree level.
\end{itemize}

The last point deserves a couple of comments. The first is about the sign of the soft terms. The sign of sfermion masses is associated to their $\SU{2}$ isospin along the $\Gen{3} = \sigma_3/2$ direction. The light $f_\alpha$ fields have $\Gen{3} = -1/2$ and get positive soft masses, while the heavy $\phi_\alpha$ have $\gen{3} = 1/2$ and get negative soft masses (the $\overline{\phi}$ fields are $\SU{2}$ singlets and have zero soft mass at the tree level). The fact that the $\phi_\alpha$ fields get negative soft mass is not worrisome, as the leading contribution to their mass is the supersymmetric term $M_\alpha \gg \tilde m$ in \eq{M}. On the contrary, negative soft masses for the light fields would have lead to a lethal spontaneous breaking of color and electric charge. The welcome positiveness of light sfermion soft masses, on the other hand, was not a priori guaranteed. It therefore reinforces the internal consistency of the model.

The second comment is about the supertrace constraint~\cite{Ferrara:1979wa}. The model we are considering is non-anomalous. The supertrace constraint then implies that the sum of all supersymmetry breaking sfermion masses vanishes at the tree level. As a consequence, the positiveness of the MSSM sfermion masses forces some sfermions with the same $\GSM$ quantum numbers~\cite{Dimopoulos:1981zb} to have a negative soft mass. This has been often considered to be an obstacle to generating sfermion masses at the tree-level in non-anomalous theories. The way out considered here is the one that goes under the name of ``tree-level gauge mediation'' (TGM), in which the sfermions with negative soft masses get a large, positive supersymmetric mass term and play the role of the chiral messengers of minimal gauge mediation. In fact, supersymmetry breaking schemes can be classified by the way they overcome the vanishing supertrace constraint, which holds at the tree level in the presence of a renormalizable K\"ahler and traceless (non-anomalous) gauge generators. In gravity mediation~\cite{Nilles:1982ik}, the supertrace does not vanish because of non-renormalizable K\"ahler, in the case of anomalous U(1)~\cite{Barbieri:1982jc} it does not vanish because the generators are not traceless, in the case of ordinary gauge mediation~\cite{Dine:1981za} it does not vanish because soft terms arise at the loop level, in the case of tree-level gauge mediation~\cite{Nardecchia:2009ew} it \emph{does} vanish and, as said, the positive soft terms of light fields is compensated by the negative soft terms of heavy fields generating gaugino masses.

Gaugino masses are generated at the one loop, as in minimal gauge mediation, because the chiral messengers $\phi_\alpha$, $\overline{\phi}_\alpha$ are coupled to supersymmetry breaking through the superpotential term in \eq{chiralmessengers} and to the MSSM gauginos through the SM gauge interactions. At the messenger scale $y_\alpha M$, gaugino masses are given by
\begin{equation}
\label{eq:gauginos}
M_i = 12\, \frac{a^2}{\sqrt{a^2-b^2}} \frac{\alpha_i}{4\pi} \frac{F}{M} ,
\end{equation}
where $\alpha_i = g^2_i/(4\pi)$ are the SM gauge constants and the form \eq{gauginos}, with the gauge coupling appropriately renormalized, is preserved by the one-loop running. In the latter approximation, the gaugino mass ratios at the weak scale are approximately $M_1 : M_2 : M_3 \sim 1 : 2 : 7$ and the gluino mass is approximately given by
\begin{equation}
\label{eq:gluino}
M_3(\text{TeV}) \approx 12 \frac{a^2}{\sqrt{a^2-b^2}} \frac{\alpha_3(\text{TeV})}{4\pi} \frac{F}{M} \approx 0.35\,  \tilde m .
\end{equation}
The previous equation shows that the ratio of gaugino and sfermion masses is fixed and is not hierarchical, despite the gaugino masses arise at the one-loop level and the sfermion masses at the tree level. The loop factor suppression of gaugino masses is compensated by two enhancements: the factor $a^2/\sqrt{a^2-b^2} \approx 5$, predicted by the 3--2 model, and the factor $12=3\times 4$ corresponding to the three vectorlike family of messengers.

The copious number of messengers charged under the SM gauge group requires a lower limit on the scale $M$ \footnote{The messenger masses are actually given by $y_\alpha M$, but for the sake of this argument, we can conservatively take $y_\alpha \sim 1$. } of the messengers in order to avoid Landau poles below the unification scale: $M \gtrsim 10^{11}\GeV$. As a consequence,
\begin{equation}
\label{eq:hbound}
h = \frac{F}{M^2} \sim \frac{\tilde m}{M} \lesssim 10^{-8} \frac{\tilde m}{\text{TeV}} .
\end{equation}
Such a bound is well in line with the assumption $h\ll 1$, which allows the model to be calculable. On the other hand, it also means that dimensional transmutation only accounts for a part of the hierarchy between the Planck and the weak scale ($\Lambda_3 \gtrsim 10^{12}\GeV$), with the remaining part accounted for by the smallness of $h$. This is quite a common situation in calculable models of dynamical supersymmetry breaking. A model with fewer messengers would allow $M$ to be lower (and $h$ to be larger) at the price of enhancing the ratio between sfermion and gaugino masses in \eq{gluino}. Another consequence of the presence of a significant number of messengers is that the $\SU{2}$ gauge coupling is IR free.

\section{Yukawa interactions and the Higgs}
\label{sec:yukawas}

In order to account for the SM fermion masses, we need to account for the MSSM Higgs doublets and the Yukawa superpotential. The latter couples two MSSM matter fields to a Higgs doublet. As the MSSM fields correspond to the $\Gen{3} = -1/2$ component of $\SU{2}$ doublets, the MSSM Higgs doublets must correspond to the $\Gen{3} = 1$ components $h^+_u, h^+_d$ of $\SU{2}$ triplets $H_u$ and $H_d$:
\begin{table}[H]
\begin{centering}
\begin{tabular}{ccc|c}
 & $\SU{3}$ & $\SU{2}$ & $\GSM$  \tabularnewline
$Q$ & $3$ & $2$ & 1 \tabularnewline
$\Uc$ & $\overline{3}$ & $1$ & 1 \tabularnewline
$\Dc$ & $\overline{3}$ & $1$ & 1 \tabularnewline
$L$ & $1$ & $\overline{2}$ & 1 \tabularnewline
 \hline
$\Phi$ & 1 & 2 & $R_\text{SM+$\nu^c$}$ \tabularnewline
$\overline{\phi}$ & 1 & 1 & $\overline{R}_\text{SM+$\nu^c$}$ \tabularnewline
$H_u$ & 1 & 3 & $(1,2,+\nicefrac[]{1}{2})$  \tabularnewline
$H_d$ & 1 & 3 & $(1,2,-\nicefrac[]{1}{2})$ \tabularnewline
\end{tabular} .
\par\end{centering}
\label{tab:all}
\end{table}
\noindent
We thus have two additional pairs of Higgs doublets $h^0_u, h^0_d$ and $h^-_u, h^-_d$ corresponding to $\Gen{3} = 0$ and $\Gen{3} = -1$ respectively. The Yukawa superpotential is in the form
\begin{equation}
\label{eq:yukawas}
W_Y = \lambda^u_{\alpha\beta} \Phi_\alpha \Phi_\beta H_u + \lambda^d_{\alpha\beta} \Phi_\alpha \Phi_\beta H_d ,
\end{equation}
where the (unique) contraction of $\SU{2}$ indexes is understood. The couplings $\lambda^{u,d}_{\alpha\beta}$ are of course non-vanishing only when the interaction term is gauge invariant,
\begin{equation}
\label{eq:yukawaexplicit}
\begin{aligned}
\lambda^u_{\alpha\beta} \Phi_\alpha \Phi_\beta H_u &\equiv \lambda^U_{ij} \Phi^{u^c}_i \Phi^q_j H_u  + \lambda^N_{ij} \Phi^{\nu^c}_i \Phi^l_j H_u, \\
\lambda^d_{\alpha\beta} \Phi_\alpha \Phi_\beta H_d &\equiv   \lambda^D_{ij} \Phi^{d^c}_i \Phi^q_j H_d +  \lambda^E_{ij} \Phi^{e^c}_i \Phi^l_j H_d .
\end{aligned}
\end{equation}

One pair of Higgs doublets, $h^+_u$ and $h^+_d$, is coupled to the MSSM fields in the Yukawa superpotential and therefore plays the role of the MSSM pair of Higgs doublets. They need to get vevs in order for the SM fermions to be generated. The Yukawa terms with $h^0_u, h^0_d$ and $h^-_u, h^-_d$ can be neglected at low energy, as they involve heavy messenger fields and can be neglected at low energy.

\section{The Higgs sector}
\label{sec:Higgs}

In order to complete the model, the Higgs interactions must be specified. Different options are available: the Higgs sector is model-dependent. We are not interested here in identifying the best possible realization of the Higgs sector, nor to solving the $\mu$-problem, we just provide an example showing that a phenomenologically viable Higgs sector can be obtained.

The simplest possibility to account for the $\mu$-term is to introduce a $\SU{2}$-invariant $\mu$-term in the superpotential,
\begin{equation}
\label{eq:WH1}
W_H = \mu H_u H_d = \mu(h^+_u h^-_d + h^0_u h^0_d + h^-_u h^+_d) .
\end{equation}
The Higgs soft terms are generated together with the sfermion masses at the scale $M$:
\begin{equation}
\label{eq:mmh}
m^2_{h^+_u} = m^2_{h^+_d} = -2 \tilde m^2, \quad
m^2_{h^0_u} = m^2_{h^0_d} = 0, \quad
m^2_{h^-_u} = m^2_{h^-_d} = 2 \tilde m^2 .
\end{equation}
Note that the soft masses of the Higgs doublets coupling to the MSSM matter fields are negative at the tree level. This is because the gauge invariance of the Yukawa interactions forces the Higgs mass terms to be given by
\begin{align*}
\label{eq:yukawainvariance}
m^2_{h^+_u} &= -(\tilde m^2_{u^c} + \tilde m^2_q) = -2\tilde m^2 < 0 \\
m^2_{h^+_d} &= -(\tilde m^2_{d^c} + \tilde m^2_q) = -2\tilde m^2 < 0 .
\end{align*}
The $B\mu$ term vanishes at the tree-level at the scale $M$, but an approximately $\SU{2}$-invariant $B\mu$ term is generated radiatively by the running in the form
\begin{equation}
\label{eq:LBmu}
\mathcal{L}_{B\mu} = m^2_{ud}\, (h^+_u h^-_d + h^0_u h^0_d + h^-_u h^+_d) .
\end{equation}

Despite $\SU{2}$ is fully broken at the scale $M$, the Higgs lagrangian below $M$ accidentally conserves $\Gen{3}$. As a consequence, the 3 pairs of Higgs doublets $h^+_u, h^-_d$, $h^0_u, h^0_d$, and $h^-_u, h^+_d$ are not mixed by any mass term. General results on such type of models are collected in the Appendix. As shown there, only one out of the three pairs of Higgs doublets, $h^+_u, h^-_d$, gets a vev, while all the other doublets, in particular $h^+_d$, have zero vev. This is not phenomenologically acceptable, as a non-vanishing vev for $h^+_d$ is necessary in order to give rise to the down quark and charged lepton masses.

In order to obtain $\VeV{}{h^+_d} \neq 0$, we need to break $\Gen{3}$ in the TeV-scale lagrangian. As for all accidental symmetries, such breaking can be provided by non-renormalizable operators. In our case, the lowest order relevant operator is in the form
\begin{equation}
\label{eq:NR}
\alpha \frac{LLH_u H_d}{\Lambda}
\end{equation}
(not to be confused with the Weinberg operator generating neutrino masses, here $L$ is the SM-singlet field of the 3--2 model), where $\Lambda > M$ is a cutoff and again $\SU{2}$ contractions are understood. After plugging the vev of $L$, the above operator gives rise to additional contributions to the $\mu$ and $B\mu$ terms:
\begin{align}
\label{eq:WH2}
W_H &= \mu H_u H_d = \mu(h^+_u h^-_d + h^0_u h^0_d + h^-_u h^+_d) + \mu_5 (h^+_u h^0_d - h^0_u h^+_d) \\
\mathcal{L}_{B\mu} &= m^2_{ud}\, (h^+_u h^-_d + h^0_u h^0_d + h^-_u h^+_d) + (m^2_{ud})_5 (h^+_u h^0_d - h^0_u h^+_d),
\end{align}
with
\begin{equation}
\label{eq:mu5}
(m^2_{ud})_5 = \frac{2 a^2}{\sqrt{c\,(a^2-b^2)}} \,\mu_5 \tilde m \approx 8.1\, \mu_5 \tilde m.
\end{equation}
We have verified numerically that in the presence of the above corrections to the $\SU{2}$-invariant $\mu$ and $B\mu$ terms, one can obtain $\VeV{}{h^+_d} \neq 0$, as desired.

A few comments are in order:
\begin{description}[leftmargin=0mm]
\item[-]\hspace*{-2mm}
The fact that the vev of $Y=-1/2$ Higgs fields is shared among $h^+_d$, $h^0_d$, and $h^-_d$, with the largest component possibly in $h^-_d$ (this is certainly the case in the limit of small $\mu_5$, $(m^2_{ud})_5$), can explain the suppression of the bottom and tau quark masses compared to the top quark mass.
\item[-]\hspace*{-2mm}
In the presence of $\mu_5\neq 0$ it could be possible to do without the $\SU{2}$-invariant $\mu$-term introduced by hand. In such a case, there would be no need to explain the presence in the superpotential of a $\ord{\text{TeV}}$ explicit mass term. Still, the (accidental) relation $\Lambda \sim \alpha (a^2-b^2) M^2/\tilde m\approx 0.075\, \alpha\, M^2/\tilde m$ would need to be invoked in order to have $\mu_5 = \alpha (a^2-b^2)\,M^2/\Lambda \sim \tilde m$. Depending on the value of $M$, the coefficient $\alpha$ in \eq{NR} could have to be small in order for $\Lambda$ not to exceed $M_\text{Pl}$.
\item[-]\hspace*{-2mm}
In the limit in which $\mu = 0$ and the $\mu$-term is provided by the operator in \eq{NR}, the $\mu$-$B\mu$ problem of gauge mediation is absent, as $B\mu/\mu\sim \tilde m$, with no loop-factor involved. However the numerical coefficient in the previous relation turns out to be largish (see \eq{mu5}).
\end{description}

\section{1-loop effects}
\label{sec:1loop}

Gaugino masses are generated at one-loop and are discussed in Section~\ref{sec:coupling}. Let us discuss here the contributions to the soft terms associated to one-loop corrections to the K\"ahler function. We will work at the first order in $F/M$. We are interested in particular to the possibility to generate $A$-terms large enough to give a non-negligible contribution to the one-loop corrections to the lightest Higgs mass.

Unlike what happens in minimal gauge mediation (see however~ \cite{Evans:2011bea}), non-vanishing $A$-terms are generated by the presence of couplings between matter and chiral messengers in the superpotential,
\begin{equation}
\label{eq:MMC}
W \supset y_{\alpha\beta} L_2 f_\alpha \overline{\phi}_\beta + \frac{\lambda^u_{\alpha\beta}}{\sqrt{2}} (f_\alpha \phi_\beta + \phi_\alpha f_\beta) h^0_u + \frac{\lambda^d_{\alpha\beta}}{\sqrt{2}} (f_\alpha \phi_\beta + \phi_\alpha f_\beta) h^0_d ,
\end{equation}
and by the gauge coupling between matter (in doublets of $\SU{2}$) and vector messengers. The couplings above give also rise to two loop contributions to the soft sfermion masses that could in principle spoil the solution of the flavour problem claimed above. We will show below that this is not the case.

The $A$-terms generated in the scalar potential $V$ by the interactions in \eq{MMC} are in the form
\begin{equation}
\label{eq:Aterms1}
V \supset A^D_{ij} \tilde d^c_i \tilde q_j h^+_d + A^U_{ij} u^c_i \tilde q_j h^+_u + A^E_{ij} \tilde e^c_i \tilde l_j h^+_d + A^N_{ij} \tilde n^c_i \tilde l_j h^+_u,
\end{equation}
with, in matrix notation,
\begin{equation}
\label{eq:Aterms2}
\begin{aligned}
A_D &= \lambda_D A_q + A_{d^c}^T \lambda_D, &
A_U &= \lambda_U A_q + A_{u^c}^T \lambda_U, \\
A_E &= \lambda_E A_l + A_{e^c}^T \lambda_E, &
A_N &= \lambda_N A_l + A_{n^c}^T \lambda_N,
\end{aligned}
\end{equation}
and
\begin{align}
\label{eq:Aterms3}
&A_q = -\frac{1}{32\pi^2} \frac{F_L}{M_L}\big(
2 y^*_q y^T_q +
\lambda^\dagger_U \lambda^{\phantom{\dagger}}_U +
\lambda^\dagger_D \lambda^{\phantom{\dagger}}_D \big) & \hspace*{-3mm}
&A_l = -\frac{1}{32\pi^2} \frac{F_L}{M_L}\big(
2 y^*_l y^T_l +
\lambda^\dagger_N \lambda^{\phantom{\dagger}}_N +
\lambda^\dagger_E \lambda^{\phantom{\dagger}}_E \big) \notag \\
&A_{d^c} = -\frac{1}{32\pi^2} \frac{F_L}{M_L}\big(
2 y^*_{d^c} y^T_{d^c} + 2 \lambda^\dagger_D \lambda^{\phantom{\dagger}}_D \big) &
&A_{e^c} = -\frac{1}{32\pi^2} \frac{F_L}{M_L}\big(
2 y^*_{e^c} y^T_{e^c} + 2 \lambda^\dagger_E \lambda^{\phantom{\dagger}}_E \big) \\
&A_{u^c} = -\frac{1}{32\pi^2} \frac{F_L}{M_L}\big(
2 y^*_{u^c} y^T_{u^c} + 2 \lambda^\dagger_U \lambda^{\phantom{\dagger}}_U \big) &
&A_{n ^c} = -\frac{1}{32\pi^2} \frac{F_L}{M_L}\big(
2 y^*_{n^c} y^T_{n^c} + 2 \lambda^\dagger_N \lambda^{\phantom{\dagger}}_N \big) . \notag
\end{align}
As \eqs{Aterms3} show, the $A$-terms turn out to be determined by the vevs of $L_1$ only, through the ratio $F_L/M_L$. This represents, as for the gauginos, a source of enhancement:  $F_L/M_L = (a^2 F)/(\sqrt{a^2-b^2}M) \approx 5 F/M$, which partially compensates the 1-loop suppression of the $A$-terms compared to the tree-level sfermion masses.

The contributions due to the vector messengers are proportional to the unknown $\SU{2}$ coupling $\gh{2}^2$. They can be suppressed ad libitum by taking $\gh{2}$ small enough (in the 3--2 model $\gh{2}$ is supposed to be perturbative) and they turn out to be small (few \% of $\tilde m$) even for $\gh{2} \sim 1$. This is due to the fact that they do not enjoy the $F_L/M_L$ enhancement and to a combination of numerical factors. We therefore neglect them in the following.

\begin{figure}
 \centering
\includegraphics[width=0.49\textwidth]{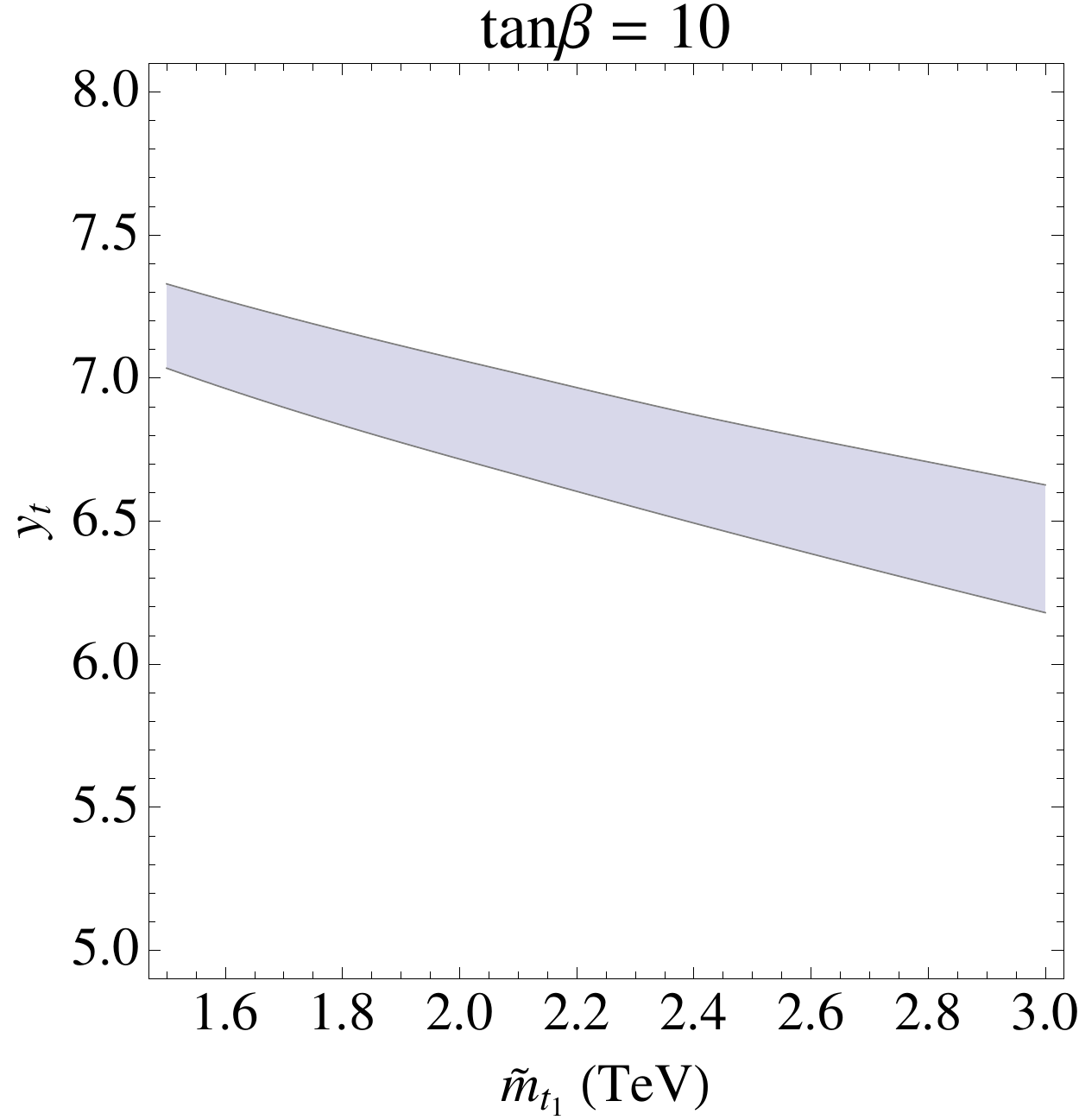} \hfill
\includegraphics[width=0.49\textwidth]{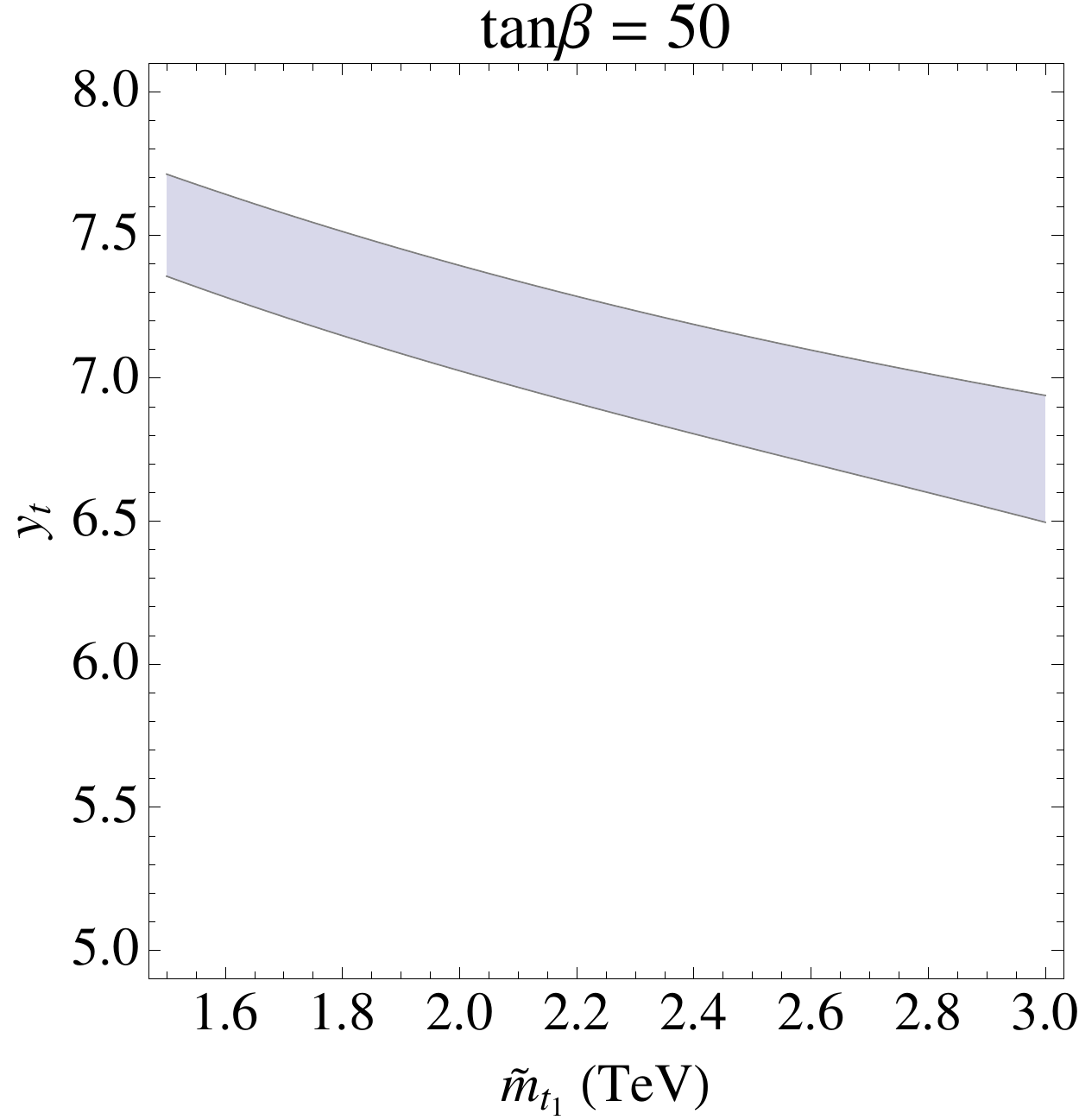}
 \caption{Range of values of $y_t$ giving a stop $A$-term large enough to account for $124\GeV < m_h < 126\GeV$, plotted as a function of the lightest stop mass $\tilde m_t$ for $\tan\beta = 10$ and $\tan\beta = 50$. The plots assume the specific realization of the model-dependent Higgs sector described in Section~\ref{sec:Higgs}.}
\label{fig:higgs}
\end{figure}

The Higgs mass is sensitive to the top $A$-term $A_t$ defined by $A^U_t = A_t\lambda_t$, where $\lambda_t$ is the top Yukawa. \Eqs{Aterms2} and~(\ref{eq:Aterms3}) give
\begin{equation}
\label{eq:Atermsquantitative}
A_t = -\frac{1}{(4\pi)^2} \frac{F_L}{M_L} \left(
y^2_{q_3} + y^2_{u^c_3} + \frac{3}{2} \lambda^2_t \right), \qquad
\fracwithdelims{|}{|}{A_t}{M_3} = \frac{y^2_{q_3} + y^2_{u^c_3} + (3/2) \lambda^2_t}{12 g^2_3}
\end{equation}
where $y^2_{q_3} = (y^*_q y^T_q)_{33}$, $y^2_{u^c_3} = (y^*_{u^c} y^T_{u^c})_{33}$. The relations above hold at the messenger scale, where the SM coupling $g_{3}$ is smaller and the yukawas are larger than their values at $M_Z$. A ratio $A_t/M_3\sim 1$ can be obtained at the messenger scale for $\lambda_t \sim y_{q_3} \sim y_{u^c_3} \sim 1.5$. Further enhancements, as required to obtain a value of the Higgs mass in the range 125--126$\GeV$ for reasonable values of $\tilde m$, require the unknown couplings $y$ to be semi-perturbative. This is shown in \Fig{higgs}, where the value of $y_t$ required to reproduce an Higgs mass in the range $124\GeV < m_h < 126\GeV$ is shown as a function of the lightest stop mass $\tilde m_t$ for two values of $\tan\beta$, 10 and 50. We have assumed for simplicity that $y_{q_3} = u_{u^c_e} \equiv y_t$. As anticipated, values of $y_t$ for which the perturbative expansion is barely valid are needed in order to reproduce the observed value of the Higgs mass. 
In order to obtain the plots in the Figure, we have used {\tt SOFTSUSY}~\cite{Allanach:2001kg}. 


\medskip

One-loop contributions to sfermion masses at $\ord{F/M}$ can also in principle arise. The contributions mediated by chiral messengers vanish because they effectively couple to one source of supersymmetry breaking only, $L_1$~\cite{Dvali:1996cu}. On the other hand, contributions mediated by vector messenger do not vanish. However, they are small, as the vector contributions to $A$-terms, and for the same reason. Analogous conclusions hold, with the superpotential we have assumed, for the 1-loop contributions to $\mu$ and $B\mu$ terms.

\section{2-loop corrections to sfermion masses}
\label{sec:2loop}

Although parametrically suppressed by a two-loop factor compared to the tree-level values, two-loop corrections to the soft masses can in principle be relevant, especially for the flavour problem.

There are three classes of contributions: the standard gauge mediation ones, the ones due to the couplings between matter and chiral messengers in \eq{MMC}, and the ones due to the coupling to vector messengers. The contributions due to the couplings to vector messengers are negligible (especially if $\gh{2}$ is relatively small). The standard gauge mediation ones can be hardly larger than $\ord{1\%}$ and are flavour blind. The ones due to the couplings in \eq{MMC} are also small enough to be ignored in the computation of sfermion masses (for $y_\alpha \sim 1$, they give a $\ord{3\%}$ correction), but they can be relevant for flavour processes. More precisely, the second and third couplings in \eq{MMC} are proportional to the MSSM Yukawas and therefore only give rise to harmless minimal flavour violating~\cite{Buras:2000dm} (MFV) contributions. The first coupling, on the other hand, is proportional to unknown Yukawas $y_\alpha$, which can in principle be largely off-diagonal in the basis in which the MSSM Yukawas are diagonal, thus providing non-MFV contributions to the soft masses. To show that the latter are also under control, let us write them, in matrix form, as follows:
\begin{equation}
\label{eq:2loops}
\delta \tilde{m}^2_f =2 \frac{y^*_f y^T_f}{(4\pi)^2} \left(\frac{T}{2(4\pi)^2}- 2c^r_f \frac{g^2_r}{(4\pi)^2}+\frac{y^*_f y^T_f}{(4\pi)^2}\right) \left(\frac{F_L}{M_L}\right)^2,
\end{equation}
where $f = q, u^c, d^c, l, n^c, e^c$,
\begin{equation}
\label{eq:T}
T = \tr \big( 6 y^{\phantom{\dagger}}_q y^\dagger_q +
3 y^{\phantom{\dagger}}_{u^c} y^\dagger_{u^c} +
3 y^{\phantom{\dagger}}_{d^c} y^\dagger_{d^c} +
2 y^{\phantom{\dagger}}_l y^\dagger_l +
y^{\phantom{\dagger}}_{n^c} y^\dagger_{n^c} +
y^{\phantom{\dagger}}_{e^c} y^\dagger_{e^c} \big),
\end{equation}
and $c^r_f$ is the quadratic Casimir of the representation $f$ with respect to the SM gauge factor $r$.

We are now in the position of studying the bounds on the off-diagonal elements of $\delta \tilde m^2$ from flavour physics. The off-diagonal elements have to be computed of course in the basis in which the mass matrix of the fermions involved in the process is diagonal. By using the bounds in~\cite{Isidori:2010kg} we find, in the squark sector
\begin{equation*}
\begin{aligned}
& \left[y^*_q y^T_q \left(T/2- 2 c^r_q g^2_r +y^*_q y^T_q \right)\right]^D_{12}, \;
\left[y^*_{d^c} y^T_{d^c} \left(T/2- 2 c^r_d g^2_r +y^*_{d^c} y^T_{d^c} \right)\right]^D_{12}  < 1.5 \text{--} 23  \\
& \left[y^*_q y^T_q \left(T/2- 2 c^r_q g^2_r +y^*_q y^T_q \right)\right]^D_{13}, \;
\left[y^*_{d^c} y^T_{d^c} \left(T/2- 2 c^r_d g^2_r +y^*_{d^c} y^T_{d^c} \right)\right]^D_{13}  <  (0.5 \text{--} 1.5)\cdot 10^2  \\
& \left[y^*_q y^T_q \left(T/2- 2 c^r_q g^2_r +y^*_q y^T_q \right)\right]^D_{23}, \;
\left[y^*_{d^c} y^T_{d^c} \left(T/2- 2 c^r_d g^2_r +y^*_{d^c} y^T_{d^c} \right)\right]^D_{23}  < (1.5 \text{--}  4.5)\cdot 10^2  \\
& \left[y^*_q y^T_q \left(T/2- 2 c^r_q g^2_r +y^*_q y^T_q \right)\right]^U_{12}, \;
\left[y^*_{u^c} y^T_{u^c} \left(T/2- 2 c^r_u g^2_r +y^*_{u^c} y^T_{u^c} \right)\right]^U_{12}  < 6 \text{--}  75
,
\end{aligned}
\end{equation*}
where $D$ and $U$ denote the bases in which the up quark and down quark mass matrices are diagonal respectively. The weaker bounds assume that only one insertion at a time is considered, with the others set to zero. The stronger ones assume that the left- and right-handed insertions are both non-vanishing and equal in size. Analogous limits can be obtained in the slepton sector. In the limit in which all yukawa are equal, $y_q = y_u = y_d = y_l = y_n = y_e \equiv y$, and neglecting the negligible (for the purpose setting the limits below) gauge contribution, we get
\begin{align*}
\left[y^* y^T \left(8 \tr(y^* y^T) +y^* y^T \right)\right]^D_{12} & < 1.5  \\
\left[y^* y^T \left(8 \tr(y^* y^T) +y^* y^T \right)\right]^D_{13} & <  0.5 \cdot 10^2 \\
\left[y^* y^T \left(8 \tr(y^* y^T) +y^* y^T \right)\right]^D_{23} & <  1.5 \cdot 10^2  \\
\left[y^* y^T \left(8 \tr(y^* y^T) +y^* y^T \right)\right]^U_{12} & <  6
.
\end{align*}
Even for anarchical yukawas with large off-diagonal entries, we see that the bounds are easily satisfied. Taking, for the sake of illustration, $\tr(y^* y^T) = 3 Y$ and $(y^* y^T)_{ij} = Y$, the bounds above are satisfied for $Y <0.2 $. This bound guarantees that the bound on the $A$-terms, which we also give for completeness, are satisfied:
\begin{align*}
(y^{\phantom{\dagger}} y)^D_{12} & < 0.5 \cdot10^2  \frac{\tilde m}{1\TeV} \frac{100\MeV}{m_s} &
(y^{\phantom{\dagger}}_q y_q)^D_{13} & < 0.5\cdot 10^3 \frac{\tilde m}{1\TeV} \frac{4\GeV}{m_b} \\
(y^{\phantom{\dagger}}_q y_q)^D_{23} & < 0.6\cdot 10^2 \frac{\tilde m}{1\TeV} \frac{4\GeV}{m_b} &
(y^{\phantom{\dagger}} y)^U_{12} & < 0.5\cdot 10^3 \frac{\tilde m}{1\TeV} \frac{1\GeV}{m_c}  .
\end{align*}
In the light of the bounds above, sizeable contributions to the Higgs mass from large $A$-terms  require a mild flavour structure in the coupling $y_{\alpha\beta}$.

\section{Summary}
\label{sec:conclusions}

We have presented a simple, complete, calculable, and phenomenologically viable model of dynamical supersymmetry breaking directly coupled to the MSSM. Supersymmetry breaking is communicated ``directly'' to the MSSM fields in the sense that the latter are directly coupled to the hidden sector $\SU{2}$ vector fields.

Supersymmetry breaking is transmitted to the MSSM fields by two sets of fields: the $\SU{2}$ vectors and the chiral $\SU{2}$ partners of the MSSM fields. The $\SU{2}$ vectors generate sfermion masses at the tree-level. Sfermion masses turn out to be universal and flavour blind, thus solving the flavour problem. The $\SU{2}$ partners of the MSSM fields behave as the chiral messengers of minimal gauge mediation and generate gaugino masses (and small corrections to the sfermion masses) at the loop level. Gaugino masses are not suppressed and are predicted to be of the same order as sfermion masses.

Sizeable contributions to the MSSM $A$-terms can arise because the chiral messengers have so-called "matter-messenger'' interactions, parameterized by free couplings. At the same time, the corresponding contributions to sfermion masses can easily made small enough in order not to spoil the solution of the flavour problem.

The Higgs sector is model-dependent. We have considered a possible implementation that predicts the existence of two additional pairs of Higgs doublets on top to the ones responsible for the SM fermion masses. As a byproduct, we have studied in the Appendix some properties of a class of models with $n$ pairs of Higgs doublets, $h^i_u$, $h^i_d$, $i=1\ldots n$.

The supersymmetry breaking model we have illustrated is an example, concrete and complete but far from unique, of the simple mechanism we used to directly communicate dynamical supersymmetry breaking to the observable fields. Several different implementations can be imagined. It is for example possible to use the U(1) factor of the 4--1 model to communicate supersymmetry breaking~\cite{FC12}.

\section*{Acknowledgements}

We are grateful to Matteo Bertolini, Gabriele Ferretti, Zohar Komargodski and David Shih for useful comments and discussions. The work of A.R.\ was supported by the ERC Advanced Grant no. 267985 ``DaMESyFla'', by the EU Marie Curie ITN ``UNILHC'' (PITN-GA-2009-23792) and the European Union FP7 ITN invisibles (Marie Curie Actions, PITN- GA-2011- 289442). Part of this work was done at the Galileo Galilei Institute for Theoretical Physics, which we thank for the kind hospitality and support.

\appendix

\section*{Appendix: electroweak symmetry breaking with $n$ pairs of Higgs doublets}
\label{sec:potential}

Let us consider a system of $n$ pairs of Higgs doublets $h^i_u$, $h^i_d$, $i=1\ldots n$. The renormalizable superpotential for such a system is just given by a generalized $\mu$-term in the form $\mu_{ij} h^i_u h^j_d$, which can always be written in a diagonal form in an appropriate flavour basis for the fields $h^i_u$ and $h^i_d$. Let us assume that the soft lagrangian is also diagonal in that basis, so that the system can be described by
\begin{equation}
\label{eq:Hlagrangian}
W = \mu_i h^i_u h^i_d, \quad
\mbox{}-\mathcal{L}_\text{soft} = m^2_{h^i_u} {h^i_u}^\dagger h^i_u + m^2_{h^i_d} {h^i_d}^\dagger h^i_d - (m^{2\,i}_{ud} h^i_u h^i_d + \text{h.c.}),
\end{equation}
where we can assume, without loss of generality, that $m^{2\,i}_{ud} \geq 0$. In this Appendix, we will study electroweak symmetry breaking in such a system. Note that, besides having an interest on its own, such a system describes the Higgs sector studied in Section~\ref{sec:Higgs}, eqs.~(\ref{eq:WH1},\ref{eq:mmh},\ref{eq:LBmu}), before introducing the $\Gen{3}$-breaking correction in \eq{NR} (i.e. for $\mu_5 = 0$, $(m^2_{ud})_5 = 0$), with the identification $(h^1_u, h^1_d) = (h^+_u,h^-_d)$, $(h^2_u, h^2_d) = (h^0_u,h^0_d)$, $(h^3_u, h^3_d) = (h^-_u,h^+_d)$.

Assuming that electric charge is not broken in the minimum, and up to SM gauge transformations, the vevs are in the form
\begin{equation}
\label{eq:vevs}
\vev{h^i_u} = e^{i\phi^i_u}
\begin{pmatrix}
0 \\ v^i_u
\end{pmatrix}, \qquad
\vev{h^i_d} = e^{i\phi^i_d}
\begin{pmatrix}
v^i_d \\ 0
\end{pmatrix} ,
\end{equation}
with $v_u^i \geq 0$, $v_d^i \geq 0$. The minimization with respect to the phases gives $e^{i\phi^i_u} e^{i\phi^i_d} = 1$. The potential can therefore be written in the form
\begin{equation}
\label{eq:VH}
V = \frac{\g^2}{2} \left(\sum_i (v^i_u)^2 - \sum_i (v^i_d)^2\right)^2 + \sum_i \Big[ m^{2}_{u,i} {v^i_u}^2 + m^{2}_{d,i} {v^i_d}^2 - 2 m^{2\,i}_{ud} v^i_u v^i_d \Big] ,
\end{equation}
with $m^{2}_{u,i} = |\mu_i|^2 + m^2_{h^i_u}$, $m^{2}_{d,i} = |\mu_i|^2 + m^2_{h^i_d}$, $\g^2 = (g^2+ {g'}^2)/4$, where $g$, $g'$ are the SM gauge couplings.

Necessary (but not sufficient) conditions for $V$ to be bounded from below are $m^{2}_{u,i} + m^{2}_{d,i}  > 2 m^{2\,i}_{ud}$ for each $i$, which we assume to be satisfied. Then it is useful to define
\begin{equation}
\label{eq:Delta0}
\Delta_0 \equiv \max \frac{1}{2\g^2} \left[
|m^2_{d,i} - m^2_{u,i}| - \sqrt{(m^2_{d,i} + m^2_{u,i})^2 - 4 (m^{2\,i}_{ud})^2}
\right]_{i=1\ldots n} .
\end{equation}
If $\Delta_0 \leq 0$, $V \geq 0 $ is bounded from below and $v^i_u = v^i_d = 0$ for each $i$ (no EWSB) is a global minimum. Let us then consider the case $\Delta_0 > 0$ and call $i_0$ the value of $i$ for which $|m^2_{d,i} - m^2_{u,i}| - \sqrt{(m^2_{d,i} + m^2_{u,i})^2 - 4 m^2_{u,i} m^2_{d,i}}$ is maximum. It is also useful to define
\begin{equation}
\label{eq:Deltapm}
\Delta^i_\pm =
\frac{1}{2\g^2} \left[
m^2_{d,i} - m^2_{u,i} \pm \sqrt{(m^2_{d,i} + m^2_{u,i})^2 - 4 (m^{2\,i}_{ud})^2}
\right] .
\end{equation}tan
Then it turns out that $V$ is bounded from below iff
\begin{equation}
\label{eq:boundness}
\Delta^i_- \leq \Delta^j_+ \quad \text{for each $i,j$.}
\end{equation}
Finally, let us assume that the condition in \eq{boundness} is satisfied, so that $V$ is bounded from below. Then (except in a vanishing measure subset of the parameter space, as discussed below) there exists a unique local minimum, coinciding with the global minimum, in which
\begin{equation}
\label{eq:minimum}
v^i_u = v^i_d = 0 \quad \text{for $i\neq i_0$}
\end{equation}
so that the potential for $v_u \equiv v^{i_0}_u$, $v_d \equiv v^{i_0}_d$ reduces to the MSSM one with (omitting the index $i_0$) $m^2_{u} + m^2_d \geq 2 m^2_{ud}$ and $m^2_u m^2_d \leq (m^2_{ud})^2$ (because of $\Delta_0 \geq 0$), so that the conditions for electroweak symmetry breaking are satisfied. The values of $\tan\beta = v_u/v_d$ and $v^2 = v^2_u + v^2_d$ are then given as usual by
\begin{equation}
\label{eq:usual}
\sin 2\beta = \frac{2 m^2_{ud}}{m^2_u+ m^2_d}, \quad
\g^2 v^2 = -\frac{m^2_d \cos^2\beta - m^2_u \sin^2\beta}{\cos^2\beta -\sin^2\beta},
\end{equation}
with $\tan\beta \gtrless 1$ if $m^2_d \gtrless m^2_u$.

\medskip

Let us now prove the relevant statements above. The proof that \eq{boundness} is a sufficient condition for $V$ to be bounded from below proceeds as follows (we assume for definitess that $m^2_d \geq m^2_u$). First we minimize $V$ with respect to $v^i_u,v^i_d$, $i\neq i_0$, for fixed $\Delta_j \equiv (v^j_u)^2- (v^j_d)^2$, $j\neq i$. The only part of the potential that is not constant and needs to be minimized is $V_i = m^{2}_{u,i} {v^i_u}^2 + m^{2}_{d,i} {v^i_d}^2 - 2 m^{2\,i}_{ud} v^i_u v^i_d$, whose minimum is given by $V^\text{min}_i = -\g^2 \Delta^i_+ \Delta_i$ for $\Delta_i \leq 0$ and by $V^\text{min}_i = -\g^2 \Delta^i_- \Delta_i$ for $\Delta_i \geq 0$. Then we observe that the condition in \eq{boundness} implies $\Delta^i_- \leq \Delta_0 \leq \Delta^i_+$, so that $V^\text{min}_i \geq -\g^2 \Delta_0 \Delta_i$ and
\begin{equation}
\label{eq:bound1}
\sum_{i\neq i_0} \Big[ m^{2}_{u,i} {v^i_u}^2 + m^{2}_{d,i} {v^i_d}^2 - 2 m^{2\,i}_{ud} v^i_u v^i_d \Big] \geq \sum_{i\neq i_0} V^\text{min}_i \geq -\g^2 \Delta_0 \Delta ,
\end{equation}
where $\Delta \equiv \sum_{i\neq i_0} \Delta_i$. Therefore,
\begin{equation}
\label{eq:bound2}
V \geq  \frac{\g^2}{2} (v^2_u - v^2_d + \Delta)^2 + m^{2}_{u} {v_u}^2 + m^{2}_{d} {v_d}^2 - 2 m^{2}_{ud} v_u v_d -\g^2 \Delta_0 \Delta  \equiv V_\text{bound}.
\end{equation}
Next, we minimize the RHS of \eq{bound2} with respect to $v_u,v_d$, which is nothing but the minimization of the MSSM potential with $m^2_u \to \tilde m^2_u = m^2_u + \g^2 \Delta$, $m^2_d \to \tilde m^2_d = m^2_d - \g^2 \Delta$. We obtain
\begin{equation}
\label{eq:Vbound}
V^\text{min}_\text{bound} = \left\{
\begin{aligned}
& -(\g^2/2) \Delta_-^2 + \g^2(\Delta_--\Delta_0)\Delta & &\text{for $\Delta \leq \Delta_-$} \\
& (\g^2/2) \Delta^2 - \g^2 \Delta_0 \Delta & &\text{for $\Delta_- \leq \Delta \leq \Delta_+$} \\
& -(\g^2/2) \Delta_+^2 + \g^2(\Delta_+-\Delta_0)\Delta & &\text{for $\Delta \leq \Delta_+$}
\end{aligned}
\right\} \geq -\frac{\g^2}{2} \Delta^2_0 ,
\end{equation}
where $\Delta_\pm \equiv \Delta_\pm^{i_0}$, which completes the proof. The proof that \eq{boundness} is a necessary condition proceeds along similar lines.

Let us now prove that if \eq{boundness} is satisfied then there the global minimum is obtained for $v^i_u = v^i_d = 0$, $i\neq i_0$, and $v_u \equiv v^{i_0}_u$, $v_d \equiv v^{i_0}_d$ minimizing the MSSM-like potential one obtains setting $v^i_u = v^i_d = 0$ for $i\neq i_0$. From the minimization in the MSSM, we know that the minimum of that potential is obtained for $v_u, v_d$ given by \eqs{usual} and the value of the potential in the minimum is given by $V_\text{min} = -(\g^2/2) \Delta_0^2$. A comparison with \eq{bound2} shows that the minimum obtained is a global one. Let us now show that there are not additional local minima. Let us denote $v^2_i = (v^i_u)^2 + (v^i_d)^2$ and $0 \leq \beta_i \leq \pi/2$ the angle such that $v^i_u = v_i \sin\beta_i$, $v^i_d = v_i \cos\beta_i$. First we show that only one $v_i\neq 0$ is allowed in a local minimum, except for the vanishing measure subset of the parameter space in which $\Delta^i_+ = \Delta^j_+$ or $\Delta^i_- = \Delta^j_-$ for some $i\neq j$. Let us consider in fact a local minimum with $v_i\neq 0$ and $v_j \neq 0$, $i\neq j$. Then
\begin{align*}
\label{eq:local1}
0 & = \frac{1}{2v^2_i} \left(\frac{\partial V}{\partial v^i_u} v^i_d + \frac{\partial V}{\partial v^i_d}v^i_u\right)  = (m^2_{u,i} + m^2_{d,i}) \sin\beta_i \cos\beta_i - m^{2\,i}_{ud} \\
0 & = \frac{1}{2v^2_i} \left( \frac{\partial V}{\partial v^i_u} v^i_u - \frac{\partial V}{\partial v^i_d} v^i_d \right) =
\g^2 (\mathbf{v_u}^2 - \mathbf{v_d}^2) + m^2_{u,i}\sin^2\beta_i - m^2_{d,i}\cos^2\beta_i ,
\end{align*}
where $\mathbf{v_u} \equiv \sum_i (v^i_u)^2$, $\mathbf{v_d} \equiv \sum_i (v^i_d)^2$. The same holds for $i\to j$. By substituting the value of $\beta_i$ one obtains from the first equation in the second one, we get
\[
\Delta^i_\pm = \g^2 (\mathbf{v_u}^2 - \mathbf{v_d}^2) = \Delta^j_\pm ,
\]
which belongs to the aforementioned vanishing measure subset of the parameter space. We can then consider the $n$ candidate local minima with $v_i\neq 0$, $i=1\ldots n$ in turn. By computing the Hessian in the minimum and using the conditions in \eq{boundness} one can then show that only the deepest critical point, the one for $i = i_0$, is a local (and global) minimum.



\providecommand{\href}[2]{#2}\begingroup\raggedright\endgroup

\end{document}